\documentclass[11pt,a4paper,english]{article}
\usepackage[T1]{fontenc}
\usepackage[utf8]{inputenc}
\usepackage{babel}
\usepackage{blindtext}
\usepackage{mathtools}
\usepackage{secdot}
\usepackage[title]{appendix}
\usepackage{amsfonts}
\usepackage{graphicx}
\usepackage{float}
\usepackage[section]{placeins}
\usepackage{booktabs}
\usepackage{subfig}
\usepackage[margin=25.4mm]{geometry}
\usepackage[labelfont=bf]{caption}
\usepackage{array}
\usepackage{csquotes}

\usepackage[backend=biber]{biblatex}
\bibliography{bibliography.bib}

\usepackage{hyperref}
\hypersetup{
    colorlinks=true,
    linkcolor=blue,
    filecolor=blue,      
    urlcolor=blue,
    citecolor=blue,
}

\title{A Decade of Evidence of Trend Following Investing in Cryptocurrencies}
\author{%
    Evans Rozario \\
    \small \href{https://www.cam.ac.uk}{University of Cambridge}\\
    \and
    Samuel Holt \\
    \small \href{mailto:samuel.holt.direct@gmail.com}{samuel.holt.direct@gmail.com}
    \and
    James West\\
    \small \href{https://globedx.com}{Globe Research}\\
    \and
    Shaun Ng\\
    \small \href{https://globedx.com}{Globe Research}
}
 
\begin{document}
\maketitle
\begin{abstract} \noindent 
Cryptocurrency markets have many of the characteristics of 20th century commodities markets, making them an attractive candidate for trend following strategies. We present a decade of evidence from the infancy of bitcoin, showcasing the potential investor returns in cryptocurrency trend following, 255\% walkforward annualised returns. We find that cryptocurrencies offer similar returns characteristics to commodities with similar risk-adjusted returns, and strong bear market diversification against traditional equities. Code available at \url{https://github.com/Globe-Research/bittrends}.
\end{abstract}

\section{Introduction}
Trend following is one of the highest capacity investment strategies of the last hundred years, with the managed futures industry managing \$325bn of investor money \cite{bookref}.
Returns in commodities markets have become less impressive over the last decade \cite{commodity} with under-performance against the S\&P 500 and poor returns \cite{commod_return}. In spite of this, trend following continues to offer low realized correlations with other traditional asset classes and thus effective diversification particularly in times of crisis \cite{aqrcentury}. 

Bitcoin \cite{btc} was introduced in 2009 as a digital currency and alternative to fiat currencies (e.g. USD, GBP, JPY) offering fast settlement, decentralization and inflationary hedge \cite{btc_vol}, resembling a commodity. In fact the US, the CFTC classifies bitcoin as a commodity \cite{pwc}. Bitcoin bears many similarities to gold: speculative safe-haven assets \cite{btc_gold} and their decentralised nature shields them from many financial variables, such as inflation and political factors. This suggests Bitcoin has effective hedging and diversification benefits \cite{gold_hedge} against global indices, but unlike gold, the returns and volatility of Bitcoin have historically been greater, leading to larger price swings and risk. With average assets under management of crypto hedge funds increasing from 21.9 million USD to 44 million USD in 2019 \cite{pwc} and the most common strategy among crypto-funds being quantitative, there is great opportunity for trend following funds to enter bitcoin. We obtain a bottom-up estimate of \$323m for maximum AUM capacity, showing significant room for growth and investment. 

To begin, we implement vanilla trend following, optimise parameters to maximise Sharpe ratio and determine the fit to digital asset markets by assessing returns, Sharpe ratio, Sortino ratio, exposure and drawdown. Utilising the Sharpe and Sortino ratios, we will gauge the attractiveness of risk-adjusted returns and upon assessing correlations, investigate the suitability of BTCUSD as a hedge against the S\&P 500. Then we extend to exponential and double exponential moving average strategies, determining features that are relevant and compare performance to the vanilla strategy and the S\&P 500. 

\section{Methodology}

\subsection*{Price data}
We use bitcoin to dollar exchange spot price data bitcoincharts \cite{btcchart} for the Bitstamp exchange, between September 13th 2011 and December 12th 2019. The data was then resampled in hourly time frames, with missing data filled forward. For this period, the data set contained 72,299 rows, of which 5,835 contained missing values and the mean date of the missing values was March 14, 2012. We expect these missing values to have resulted from the lack of trading liquidity during the infancy of Bitcoin \cite{liquid}. In this case, filling does not affect results because we are more interested in recent data and expect phenomena during 2011-2012 to be irrelevant in recent data and trends. We obtained daily OHLC data from Yahoo! Finance \cite{yfin} over the same period. 


\begin{figure}[H]
\begin{tabular}{cc}
  \includegraphics[width=0.5\textwidth]{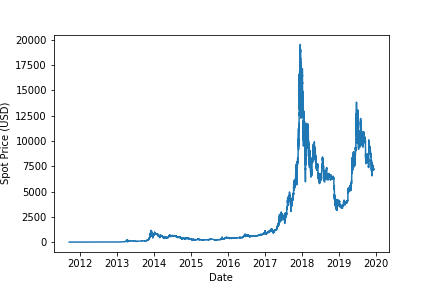} &   \includegraphics[width=0.5\textwidth]{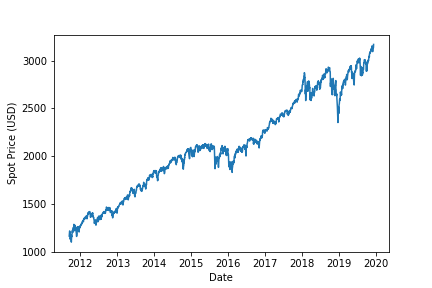} \\
(a) BTCUSD closing prices & (b) S\&P 500 closing prices
\end{tabular}
\caption{Spot price between Sept 13, 2011 and Dec 12, 2019}
\end{figure}

\subsection*{Strategy implementation}
We implemented the trend following strategies, of simple, exponential and double exponential moving average strategies. The strategy parameters, the long and short windows vary from $1-1000$ hours for BTCUSD and $1-50$ days for S\&P 500. The strategy generates a "buy" signal when the short window rolling 
average rises above the long window rolling average and similarly it generates a "sell" signal when the short window rolling average falls below the long window rolling average.

The strategies were back tested using a starting amount of 10,000 USD, and the maximal volume of respective assets were bought or sold for each "buy" or "sell" signal. We optimized for Sharpe ratio with the short and long window sizes, noting the Sortino ratio, drawdown, exposure and return for that back test. We assumed negligible transaction fees, bid-offer spread, slippage and market impact from trades.

\subsection*{Vanilla Trend Following}
We use the traditional trend estimation techniques spanning simple moving averages (SMA), exponential moving averages (EMA) and double exponential moving averages (DEMA), which we briefly summarize. 

The simple average \cite{sma} for a short and long rolling window of fixed size: let $N$ be a fixed window size and $(x_1,x_2 \dots)$ be data-points. Then for $n \geq N$ the simple average of a rolling window $(x_{n-N+1}, \dots x_{n})$ is defined as

\begin{equation}
     \mbox{SMA}(n) =  \frac{1}{N} \sum_{i = n-N+1}^{n} x_{i}
\end{equation}

\subsubsection*{Exponential Moving Average (EMA)}    
This is a weighted moving average placing greater weights on more recent prices \cite{ema}, whereas the simple average places equal weights on all prices and does not account for recent price action. Let $\alpha = \frac{2}{N+1}$ and $Y(n)$ be the closing price of the $n^{th}$ day, then the $n^{th}$ EMA is thus 

\begin{equation}
\mbox{EMA}(n) = 
\begin{dcases}
    \mbox{SMA}(1)&  n=1\\
    \alpha Y(n) + (1-\alpha)\mbox{EMA}(n-1)&  n>1
\end{dcases}
\end{equation}

\subsubsection*{Double Exponential Moving Average (DEMA)}    

Based on the EMA, the DEMA reduces noise and the lag time of signals \cite{dema}. Therefore, it is faster reacting and more receptive to fluctuations in recent prices. To compute this, we start by calculating EMA(EMA)(n), which is the $n^{th}$ EMA of the array of EMAs of our data-set. Then DEMA(n) is defined $ \mbox{DEMA}(n) = 2 \mbox{EMA}(n) - \mbox{EMA(EMA)}(n) $

\subsection*{Parameter optimization}
\begin{figure}
\begin{tabular}{cc}
  \includegraphics[width=0.5\textwidth]{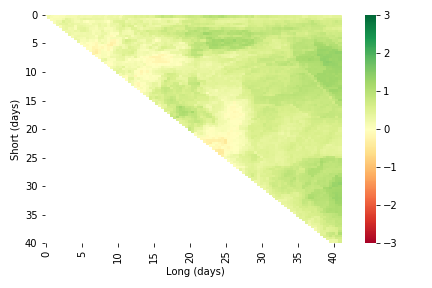} &   \includegraphics[width=0.5\textwidth]{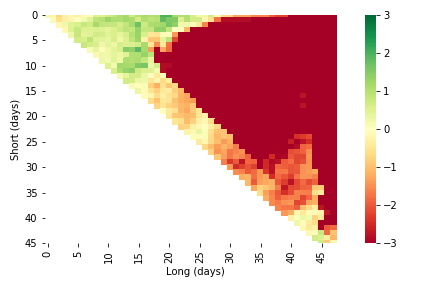} \\
(a) BTCUSD spot Sharpe ratio & (b) S\&P 500 futures Sharpe ratio
\end{tabular}
\caption{
    Risk adjusted SMA trend following returns (Sharpe ratios) in 2019. Note that BTCUSD spot is computed with hourly resolution and S\&P 500 uses daily return data. 
}
\label{fig:RASMAR}
\end{figure}

Trend following strategies have two parameters to optimize over: the long average duration and the shorter average duration. We illustrate the risk-adjusted returns surfaces according to these parameter values in Figure ~\ref{fig:RASMAR}, illustrating for 2019 the rather different return characteristics between BTCUSD and S\&P 500 in this time period. 

We employ a walk-forward approach to parameter estimation without great care for the robustness of the fitted parameter values. In particular, we take the optimal (long, short) parameter pair for the next time period to be the optimal values found in the previous period. 

\section{Results}

\subsection{Superior risk adjusted returns}
\begin{figure}[H]
\begin{tabular}{cc}
    \includegraphics[width=0.45\textwidth]{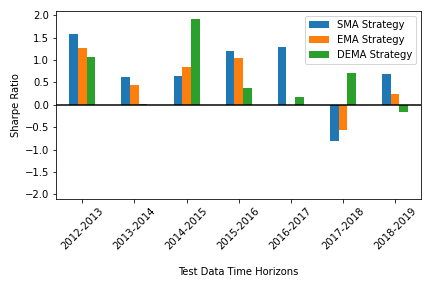} & \includegraphics[width=0.45\textwidth]{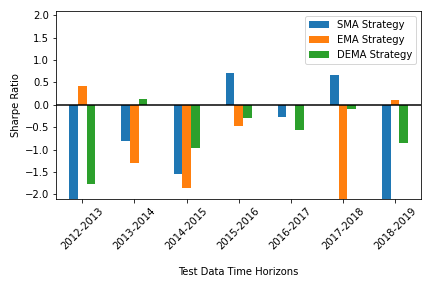} \\
    (a) BTCUSD trend following returns & (b) S\&P500 trend following returns \\
    \includegraphics[width=0.45\textwidth]{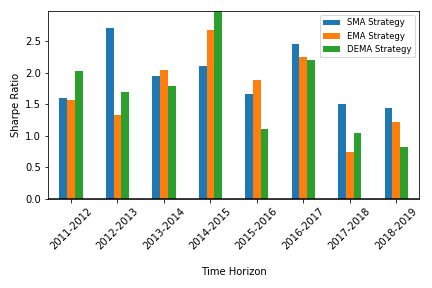} & \includegraphics[width=0.45\textwidth]{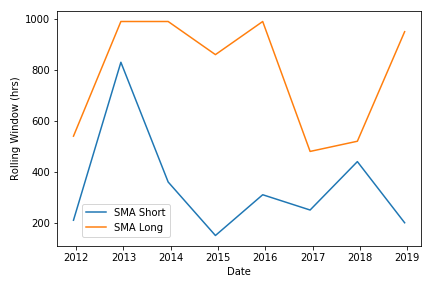} \\
    (c) BTCUSD best possible Sharpe Ratios & (d) BTCUSD parameter values \\
\end{tabular}
\caption{
    Risk adjusted returns (Sharpe ratios) from trend following the BTCUSD spot markets and the S\&P500 between 2011 and 2020. Best possible values are computed across all possible BTCUSD strategies, fitted on a rolling yearly horizon in a walkforward fashion (i.e. the parameter used for year $Y+1$ takes the optimal parameter value obtained in year $Y$.
}
\label{fig:walkforward}
\end{figure}

Over the last few years years, the SMA strategy has potential to provide attractive Sharpe ratios and risk-adjusted returns if the long and short windows are hyper-optimised. Compared to its counterparts, the optimal SMA strategy provides the best Sharpe ratios, with notably poor Sharpe ratios from EMA and DEMA in the 2017-2018 and 2018-2019 data slices. Investigating the 1 year data slices before 2016, we find impressive optimised Sharpe ratios from EMA and DEMA, often outperforming SMA but we find the optimised parameters for such Sharpe ratios are large and exceed hourly time scales. However, performing the same analysis (Figure ~\ref{fig:walkforward}) on the S\&P 500 1 year data slices, we note very low and negative Sharpe ratios, which suggests the maxima from the heat-map of S\&P 500 SMA 1 year horizon (Figure ~\ref{fig:RASMAR} (b)) are non-robust and unstable; small perturbations around maxima cause drastic shifts in Sharpe ratio.

Investigating Figure ~\ref{fig:walkforward} (b), we note that for the EMA strategy that uses the 2015-2016 data slice as training, there is no Sharpe ratio. This is because optimising long and short windows for 2015-2016 and testing on 2016-2017 with such parameters, leads to no executed trades. Using recent data slices (2015-2016, 2016-2017, 2017-2018) we note variable and sub-par performance from all strategies with Sharpe ratios < 1 and negative Sharpe ratios from SMA and EMA using 2016-2017 and DEMA using 2017-2018 as training data. Considering all training data slices, we find no predictable and attractive Sharpe ratios from using trend following strategies on 1 year training data commencing year Y, for test data commencing year Y+1. This statement should be tested again with more mature Bitcoin markets, where we can investigate with more data slices. Combining walk forward returns from BTCUSD SMA from 2011 - 2019 we achieve a return of 73700\%, which is an annualised return of 255\%.

\subsection{Diversification from traditional assets}
Over the last decade, there have been long periods with negative or no correlation between BTCUSD and S\&P 500, suggesting BTCUSD could be an effective temporal hedge against the S\&P 500 and trend following as an asset is a stock market diversifier. 

During our entire dataset (2011 - 2019) BTCUSD spot prices and equities returns exhibited a small negative correlation, albeit not statistically significant. 

Analysing Figure ~\ref{fig:walkforward} (d) and repeating the same analysis for EMA and DEMA, we notice the optimal short window is generally above 200 hrs. Moreover, for EMA and DEMA, the size of long and short window appear to be positively correlated, with correlations $>0.65$. For all strategies, window sizes show little trend with date with large variations in both long and short windows from 200-1000 hrs. However, the sample size of rolling windows is very small and given the young nature of Bitcoin, this assertion should be retested when Bitcoin markets have matured. 

\begin{figure}[H]
\begin{tabular}{cc}
  \includegraphics[width=0.5\textwidth]{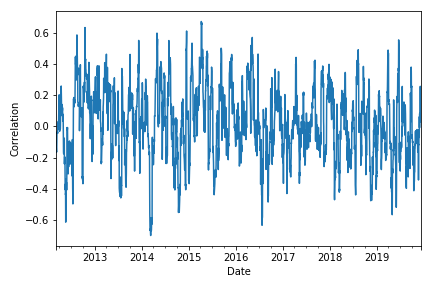} &   \includegraphics[width=0.5\textwidth]{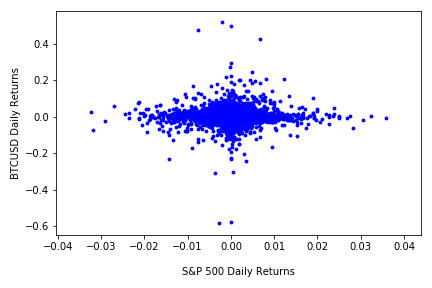} \\
(a) Daily returns correlation & (b) Daily returns scatter plot
\end{tabular}
\caption{For period 12.12.2011 - 12.12.2019: (a) 20 day rolling correlation of S\&P 500 SMA vs BTCUSD SMA daily returns data.}
\end{figure}     

\subsection{Robustness to strategy}
Throughout the decade or so long period for which we have data, we find trend following strategies to be consistently profitable regardless of averaging strategy used. Moreover the highest possible achievable risk-adjusted rewards are similar to commodities markets.

\begin{table}[h!]
    \centering
    \resizebox{0.7\textwidth}{!}{\begin{tabular}{ lccccccc }
    \toprule
    \emph{Strategy} & \emph{Short (hrs)} & \emph{Long (hrs)}
              & \emph{Sharpe Ratio} & \emph{Sortino Ratio} \\
    \midrule
    SMA         & 141 & 781 & 1.0907 & 0.7969 \\
    EMA         & 721 & 951 & 1.3515 & 12.9827 \\
    DEMA         & 791 & 981 & 1.3165 & 2.2121 \\
    
    \bottomrule
    \end{tabular}}
    \caption{\label{btcall} Best possible BTCUSD spot trend following over full period (September 13 2011 to December 12, 2019)}
    \label{table:results1}
\end{table}

\section{Conclusion}

The last decade has seen consistent and attractive returns from trend following in cryptocurrency markets. Whilst this paper has only studied the bitcoin-dollar exchange rates, the extremely high levels of correlation between cryptocurrencies in recent years means that most results will carry over to other digital assets. 

Returns are overall reasonably robust to methodology used for determining trends and to parameter choices for trend following algorithms. Over a decade long walkforward test, using a simple yearly regime for parameter estimation, we find reasonable consistency in estimates, and find that simple strategies taking simple moving averages of approximately 10 and 40 days consistently perform well, with a slight dropoff in more recent times. Sharpe ratios for such strategies are consistently around the 0.5 to 1.5 range (with only one year loss making) as with commodities trend following strategies decades ago, in spite of the different volatility characteristics of cryptocurrency markets. Of particular interest to practitioners is the notable absence of profitable intra-day trend following strategies for BTCUSD spot markets, in spite of the considerable interest afforded to such strategies. 

The last decade has been a length bull run for US equities markets, meaning that there is no data since the infancy of bitcoin covering its mechanics in a prolonged recession. Regardless, during this period there was no significant correlation between daily returns of BTCUSD exchange rates and US equities. This, combined with the strong positive returns from trend following points to cryptocurrencies as a fertile ground for portfolio diversification, particularly against the inflationary environment of 2020.

The code and results for this project can be found at: \url{https://github.com/Globe-Research/bittrends}.

\section{Acknowledgements}

The research in this paper was made possible by resources provided by Globe Research as part of Globe, a pioneering cryptocurrency derivatives exchange, available at \url{https://globedx.com}.

\newpage

\printbibliography

\newpage

\begin{appendices}

\section{Risk adjusted return surfaces}

\begin{figure}[htbp]
\begin{tabular}{cc}
  \includegraphics[width=0.5\textwidth]{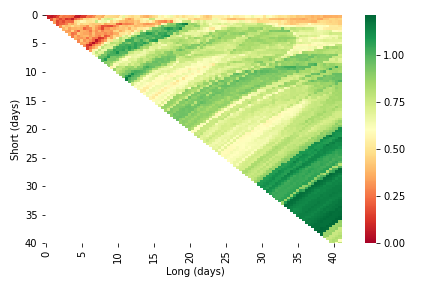} &   \includegraphics[width=0.5\textwidth]{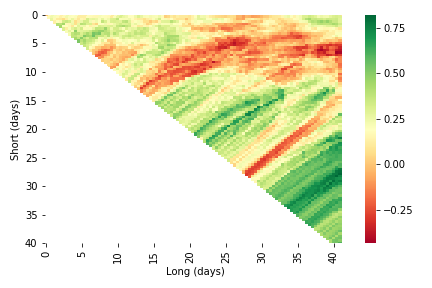} \\
(a) EMA & (b) DEMA
\end{tabular}
\caption{BTCUSD Sharpe ratio heat-maps for 2019}
\end{figure}
    
\begin{figure}[htbp]
\begin{tabular}{cc}
  \includegraphics[width=0.5\textwidth]{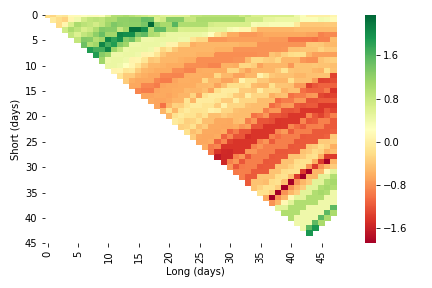} &   \includegraphics[width=0.5\textwidth]{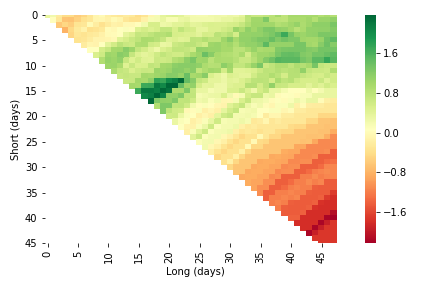} \\
(a) EMA & (b) DEMA
\end{tabular}
\caption{S\&P 500 Sharpe ratio heat-maps in 2019}
\end{figure}    

\end{appendices}

\end{document}